# Solvated Electrons and Hydroxyl Radicals at the Plasma-Liquid Interface

**Seungjun Lee [1], Hyung-Gu Kang [1], Minkwan Kim[2] and Gunsu Yun [1,3*]**
[1] Division of Advanced Nuclear Engineering, Pohang University of Science and Technology, Pohang 37673, Republic of Korea
[2] Astronautics Research Group, Faculty of Engineering and Environment, University of Southampton, Southampton SO17 1BJ, United Kingdom
[3] Department of Physics, Pohang University of Science and Technology, Pohang 37673, Republic of Korea
*E-mail: gunsu@postech.ac.kr

**ABSTRACT**

While hydroxyl radicals (·OH) play an important role as potent oxidizing agents in various plasma applications, their high reactivity confines them to a thin layer at the plasma-liquid interface, posing challenges in comprehending the intricate generation and transport processes. Similarly, solvated electrons ($e_{aq}$), highly reactive reducing agents, are expected to exhibit distribution beneath the liquid surface and interact with ·OH in the thin layer. In this study, we have determined the penetration depth and concentration of $e_{aq}$ at the interface between an atmospheric argon plasma plume and an electrolyte anode via a lock-in amplification absorbance measurement. With bias voltages from 1000 to 2500 V, the penetration depth remains approximately 10 nm, and the peak concentration near the surface reaches 1 mM. Diffusion is the primary mechanism for ·OH generation in the electrolyte, with most ·OH reacting with $e_{aq}$ at the interface, thus influencing the $e_{aq}$ distribution. In contrast, the electrolyte cathode significantly boosts ·OH generation, leading to rapid recombination into hydrogen peroxide.

**INTRODUCTION**

Plasma-liquid interactions present a unique combination of physics, chemistry, and materials science, distinct from solid-liquid or gas-liquid interactions, emerging as an important topic in plasma science and engineering [1-5]. When plasma interacts with liquid, it introduces reactive species, such as hydroxyl radicals (·OH), atomic oxygen (O·), and hydrogen peroxide ($H_2O_2$), into liquids through various processes such as sputtering, electrolysis, and photolysis [5-7]. This interaction can alter the liquid's pH, conductivity, and ionic composition. The ensuing changes in the liquid, in turn, influence plasma-induced chemical reactions or discharges, establishing a complex feedback loop and interface characteristics that make understanding the physical and chemical mechanisms challenging.

Acknowledging the activation of liquids upon contact with plasma has led to an emphasis on various applied research studies [5]. Particularly, ·OH, known for its high oxidation potential of 2.8V and high secondary reaction coefficients, can oxidize a wide range of organic and inorganic compounds [8-10]. Due to its ability to avoid generating secondary pollutants, it is recognized as one of the key radical species in applied research [10]. In advanced oxidation processes, the atmospheric pressure plasma stands out for its capability to generate abundant comparable to conventional methods like $O_3/H_2O_2$ and $H_2O_2$/UV [10-12]. Plasma-derived ·OH has been actively used in biological applications, showing promising results in wound healing [13], teeth whitening [14], apoptosis induction in tumor cells [15], fibroblast proliferation [16], angiogenesis [16], and sterilization [17]. Consequently, understanding the mechanisms behind the generation of ·OH has become increasingly important in applied research.

Solvated electron ($e_{aq}$), which is electron bound to the solution, have recently been measured through optical absorption at the plasma-liquid interface [18]. Similar to ·OH, $e_{aq}$ are considered to exist with penetration depths on the order of tens of nanometers and concentrations in the mM range at the plasma-liquid interface owing to its high reactivity [19]. The distribution of $e_{aq}$ confined at the interface implies that they reduce most radicals or oxidizing species produced by plasma. Our previous research showed that a significantly higher concentration of ·OH is obtained when the plasma is coupled directly to the liquid cathode and anode, compared to when plasma interacts with the liquid remotely [20]. This indicates that the coupling between plasma and liquid plays a more important role in generating ·OH than the mutually scavenging reaction between ·OH and $e_{aq}$, highlighting the need for simultaneous measurements of both ·OH and $e_{aq}$ for additional insights.

In the present study, we measure the absorbance of $e_{aq}$ and the concentration of ·OH at the plasma-liquid interface of an electrolyte anode. Using total internal reflection absorption spectroscopy (TIRAS), based on the approach by Rumbach *et al* [18], the absorbance of the interfacial layer is measured for a range of bias voltages applied to the electrolyte. The penetration depth and peak concentration of $e_{aq}$ are estimated from the measured absorbance using Rumbach's model. For the bias voltage range of 1000–2500 V, the average penetration depth and peak concentration of $e_{aq}$ are around 10 nm and 1 mM, respectively. The distribution of ·OH is similar to that of $e_{aq}$, with nm-scale penetration depths and mM concentrations, primarily generated through diffusion and predominantly reacting with $e_{aq}$ at the interface. By contrast, in the case of electrolyte cathode, ·OH generation is significantly higher, leading to the formation of 150 mM hydrogen peroxide ($H_2O_2$) via ·OH recombination—a level not observed in the electrolyte anode case.

The following sections will explore these findings further. We begin with a detailed methodology section outlining the TIRAS setup and ·OH measurement procedure. Next, we present the results on $e_{aq}$ and ·OH distributions and analyze the influence of bias polarity on the production and reactivity of these species. Finally, we discuss the mechanism of ·OH generation in plasma-liquid interface chemistry.

**METHOD**

1. Plasma Source and Electrolyte

The overall experimental setup is illustrated in Figure 1 (a). The atmospheric pressure plasma generated at a stainless-steel needle electrode is positioned within 1 mm above the liquid surface, and a platinum counter electrode is immersed in the electrolyte. The switching circuit, consisting of an optical coupler (TOSHIBA, TLP250) and a high-voltage IGBT (IXYS, IXGF20N300), is connected to the power supply (Stanford Research Systems, PS325). This configuration enables the application of modulated DC bias up to 2.5 kV at 25 kHz to the electrolyte, initiating a plasma discharge between the solution surface and the needle. The reaction takes place in a quartz cuvette with the internal dimensions of 5 × 5 × 5 cm, containing 30 ml of electrolyte. To control the surrounding gas environment where plasma is generated, the reactor is sealed with a cap, and high-purity argon gas is continuously purged at 100 standard cubic centimeters per minute (sccm). Bias voltage and current are measured using a high-voltage probe (Tektronix, P9015A). For plasma discharge at low voltages, the plasma is initially ignited at 2.5 kV, and then the bias voltage is reduced to the set voltage (minimum of 500 V).

The electrolyte solution is a terephthalic acid (TA) solution composed of 2 mM TA (Sigma-Aldrich Co., product number 185361) and 5 mM NaOH (Samchun Inc., Catalog Number S2501). TA is a well-known ·OH scavenger and does not react with other reactive species such as $O_2^-$, $HO_2$, and $H_2O_2$ [21, 22]. As shown in Figure 1 (b), ·OH and TA react to produce 2-hydroxy-terephthalate ion (HTA), whose concentration can be measured through fluorescence. NaOH is used to facilitate the dissolution of TA in the solution and adjust the electrolyte's conductivity without reacting with $e_{aq}$. The pH and conductivity of the TA solution are 10.1 and 520 µS/cm, respectively.

2. Measurement of Solvated Electrons and Hydroxyl Radicals

$e_{aq}$ at the plasma-liquid interface is measured by the aforementioned TIRAS method [18]. A HeNe laser beam (Thorlabs Inc., HRS015B) with a wavelength of 633 nm is directed to the central region of the plasma-liquid interface at an angle of 18° (at which total internal reflection occurs), and the reflected light is received by a photodetector (Thorlabs Inc., PDA100A2). The argon gas flow rate and the distance between the electrode and the liquid surface are carefully adjusted so that there is no distortion in the liquid surface due to the gas flow. Although the reflected laser beam expands elliptically, the expansion is small enough for all light to be captured by the photodetector. Additionally, during plasma discharge with TA solution, a slight precipitation of TA occurs at the counter electrode that is immersed in the liquid, which does not affect the laser beam path. The minute variations in the photodetector signals with the bias voltage modulation are resolved by a lock-in amplifier (Zurich Instruments, MFLI). The measured absorbance gives the estimates of the penetration depth and concentration distribution of $e_{aq}$ using the model developed by Rumbach et al [18]:

$$\frac{\Delta I}{I_0} = \frac{2\varepsilon l}{\sin\theta} \frac{\sqrt{i_{high}} - \sqrt{i_{low}}}{\sqrt{\pi \sigma_p^2 l k_2 q N_A}} \left[\left(1 + \frac{\sigma_x^2}{2\sigma_p^2}\right)\left(1 + \frac{\sigma_y^2}{2\sigma_p^2}\right)\right]^{-\frac{1}{2}} \frac{1}{\sqrt{2}} c_f \quad (1)$$

$$[e_{aq}^-](x,y) = \sqrt{\frac{i}{\pi \sigma_p^2 l k_2 q N_A}} \exp\left(-\frac{x^2}{2\sigma_p^2} - \frac{y^2}{2\sigma_p^2}\right) \quad (2)$$

where $\Delta I/I_0$ is the absorbance due to the presence of $e_{aq}$, $l$ is the average penetration depth, $\varepsilon$ is the extinction coefficient ($\varepsilon = 1.6 \times 10^6 \, M^{-1} m^{-1}$ at 633 nm) [23, 24], $\theta$ is the angle of laser incidence, $q$ is the elementary charge unit (negative of the electron charge), $N_A$ is Avogadro's number, $i$ is the time-averaged plasma current, and $i_{high}$ and $i_{low}$ are the minimum and maximum levels of the plasma current, respectively. $k_2$ is the rate constant for the second-order recombination at room temperature [24]: $2H_2O + 2(e_{aq}) \rightarrow 2(OH^-) + (H_2)_{aq}$, $2k_2 = 1.1 \times 10^{10} \, M^{-1} s^{-1}$ [24]. $\sigma_x$ and $\sigma_y$ are the radii of the laser's Gaussian intensity distribution in the x and y directions beneath the liquid surface, and $\sigma_p$ is the radius of the plasma above the liquid surface, which is determined by Gaussian fitting of the plasma image photographed by a DSLR camera (Canon, EOS 5d Mark2). The factor $1/\sqrt{2}$ in equation (1) corresponds to taking the root mean square of the lock-in amplifier measurements, and the parameter $c_f$ accounts for the peak amplitude of the first Fourier mode of the plasma current. For instance, if the current is modulated as a sine wave, $c_f$ is 1, and if modulated as a square wave with a 50% duty cycle, $c_f$ is 4/π.

The concentration of ·OH is measured based on the concentration of HTA (Sigma-Aldrich Co., product number 752525), as obtained from a calibration curve. The HTA concentration is determined by transmitting UV light at a central wavelength of 310 nm (Ocean Insight, LSM-310A LED) through a 30 ml TA solution treated with plasma for 30 minutes. Fluorescence at 425 nm is collected in the vertical direction using a compact spectrometer (RGB Photonics, Qmini). Note that TA solution is not affected by electrolysis, and the pH effect on fluorescence intensity is less than 2% when the pH level is above 4 [25].

## RESULTS & DISCUSSION

To determine the $c_f$ value in equation (1), we measure the plasma current waveform modulated by the IGBT gating as shown in Figure 2. The plasma discharge rises rapidly and decays after the gate signal, which may be approximated as a reverse sawtooth wave. The corresponding value of $c_f$ is approximately $2/\pi$.

Figure 3 shows the brightness profile of the plasma measured by a DSLR camera for the bias voltages from 500V to 2500V in 500V intervals. As the bias voltage increases, the plasma expands radially (x direction). Gaussian fitting indicates that $\sigma_p$ increases from 0.25 mm to 0.66 mm; however, the current density remains constant in the range of 5 to 6 A/m². Note that for high bias voltages, the profile somewhat deviates from the Gaussian shape, exhibiting a plateau region. This may be attributed to a pattern formation at the electrolyte anode similar to the observation by Rumbach et al [26], where a transition occurred from a uniform plasma to a ring-patterned plasma as the plasma current increases.

Diode lasers are additionally used to measure the absorbance at different wavelengths. An aspheric lens is used to focus the laser beam with the spot radius from 63 to 100 µm at the location where plasma is generated. The laser beam passes through an iris to reduce ambient light and its profile is measured at the focal length using a CCD (Thorlabs Inc., DCC1545M). The beam profile and absorbance for each wavelength are shown in Figure 4.

Figure 5 shows the measured absorbances for the bias voltage from 500 V to 2500 V with 10 V intervals. The lock-in amplifier system enables measurement of the minute change in the absorbance on the order of $10^{-5}$. The absorption profile, influenced by the bias voltage, can be divided into three regions, as follows: (1) in the 500–1000 V range, the amount of $e_{aq}$ increases with the growth of plasma radius and plasma current. (2) In the 1000–2000 V range, the plasma radius is sufficiently larger than the laser spot size, e.g., $\sigma_p$ = 0.360 mm > $\sigma_y$ = 0.252 mm at 1000 V. The laser spot falls within the region of uniform plasma density, and increasing the plasma size does not change the total absorbance. (3) In the 2000–2500V range, the plasma becomes a transitional state from a uniform profile to a distinct ring pattern, causing a reduction of the absorbance.

Figure 6 shows the estimated penetration depth and central concentration of $e_{aq}$ from the measured absorption data shown in Figures 3–5 using equations (1) and (2). Cases with a bias voltage of 500V or below, where the plasma current modulation is unstable, are excluded. The penetration depth and peak concentration remained approximately 10 nm and 1 mM, respectively, regardless of the bias voltage. The localized and high concentration of $e_{aq}$ suggests that the following reaction (3) occurs intensely at the plasma-liquid interface. The pH of the TA solution gradually increased during plasma treatment, rising from 10.10 to 10.26 after 5 minutes of treatment with a bias voltage of 2500 V. However, note that the change in pH of the solution is primarily due to the reaction $2H_2O + 2e_{aq} \rightarrow H_2 + 2OH^-$ rather than reaction (3).

$$e_{aq}^- + \cdot OH_{aq} \rightarrow OH_{aq}^- \qquad (3)$$

To determine the ·OH concentration within the electrolyte bulk, the HTA concentration is measured after subjecting a 30ml TA solution to a 5-minute plasma exposure, as illustrated in Figure 7. Additionally, to verify the effect of $e_{aq}$, measurements are taken under the condition in which the electrode served as the electrolyte cathode. These measurements revealed a significantly higher HTA concentration—approximately six times greater—compared to conditions where the electrode served as the electrolyte anode. This elevated ·OH concentration facilitates the formation of $H_2O_2$ via recombination reactions [29]. Consistently, the $H_2O_2$ concentration, determined via a colorimetric method using titanium(IV) oxysulfate (Sigma-Aldrich, product number 89532), reached approximately 150 mM, whereas no $H_2O_2$ is detected for the electrolyte anode case. This suggests that, in the electrolyte anode case, ·OH predominantly reacts with $e_{aq}$ via Reaction (3), forming $OH^-$.

To further analyze the ·OH, the concentration of ·OH and the generation rates of HTA are evaluated. Assuming ·OH as the primary scavenger of $e_{aq}$, the average penetration depth $l_e$ of $e_{aq}$ is given by [19]:

$$l_e = \sqrt{\frac{D_e}{k_{\cdot OH,e}[\cdot OH]}} \qquad (4),$$

The concentration $n_e$ of $e_{aq}$ is

$$n_e = \frac{j_e}{F}\sqrt{\frac{1}{D_e k_{\cdot OH,e}[\cdot OH]}} \qquad (5)$$

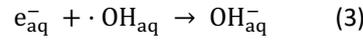

Where $D_e$=4.8 × $10^{-9}$ m²s⁻¹ is the diffusivity, $k_{\cdot OH,e}$=3.0 × $10^{10}$ M⁻¹s⁻¹ is the rate constant for reaction (3). Substituting the values for penetration depth (10 nm) and peak concentration (1 mM) into equations (4) and (5), [·OH] is approximately 1.6 mM and 1.9 mM, respectively. These values, both in the a few mM range, indicate that ·OH strongly influences the distribution of $e_{aq}$. The HTA generation rate, shown in Fig.7 (b), is approximately $dN_{HTA}/dt \approx 5×10^{-11}$ mol/s. The spatial distribution of OH concentration due to diffusion can be represented as follows:

$$n_{\cdot OH(x,y,z)} = n_{\cdot OH} \exp\left(-\frac{x^2}{2\sigma_p^2} - \frac{y^2}{2\sigma_p^2} - \frac{z}{l_{OH}}\right) \qquad (6)$$

Using $dN_{HTA}/dt = k_{TA,\cdot OH}$ [TA][·OH]V, the rate of HTA production is expressed as:

$$\frac{dN_{HTA}}{dt} = \int_0^\infty \int_{-\infty}^\infty \int_{-\infty}^\infty k_{TA,\cdot OH}[TA]n_{\cdot OH} \, exp\left(-\frac{x^2}{2\sigma_p^2} - \frac{y^2}{2\sigma_p^2} - \frac{z}{l_{OH}}\right) dxdydz = 2k_{TA,OH}[TA]n_{OH}\pi\sigma_p^2 l_{OH} \qquad (7)$$

where $k_{TA,\cdot OH}$ = 4 × $10^9$ M⁻¹s⁻¹ is the rate constant for TA-·OH reaction. Based on these calculations, we find that $n_{\cdot OH} l_{\cdot OH}$ is approximately 2.3 × 10$^{-12}$ Mm. Synthesizing these findings, we infer that ·OH exist within a concentration range of a few mM orders, penetration depths on the order of a few nm, and reacts intensely through Reaction (3), thereby influencing the distribution of e$_{aq}$.

The generation pathways of ·OH in the underwater environment are diverse. One straightforward route is the diffusion from plasma to liquid. To analyze diffusive species within the plasma, optical emission spectroscopy (OES) is used. Figure 8 shows the presence of plasma species, including ·OH and Ar. ·OH cannot accumulate beyond a certain concentration due to its high reactivity, and the maximum ·OH flux from the plasma is known to be $\Gamma_{·OH} \sim 10 \times 10^{21}$ m$^{-2}$s$^{-1}$ [19, 29]. Diffusion provides enough flux as the primary source of ·OH in the solution for the electrolyte anode case. However, the ·OH flux is insufficient to explain the high concentration of HTA in the electrolyte cathode case, implying that ·OH is generated by a mechanism other than diffusion such as ion bombardment by the cathode fall. The chemical process initiated by the bombarding cation ($\oplus$) can be described by the reaction: $\oplus + xH_2O \rightarrow xH_2O^* + H_2O^+$, followed by an ionization reaction ($nH_2O^* + H_2O^+ \rightarrow (n+1)·OH + (n+1)H^+ + ne_{aq}^-$) or a dissociation reaction ($nH_2O^* + H_2O^+ \rightarrow nH + n·OH + H_2O^+$) [27, 28].

Another possible route for ·OH generation is photolysis induced by UV radiation from the plasma. As shown in Figure 8, UV light emissions are detected below 240 nm, with photon energies exceeding the bond energy of the O-H bond of the water molecule (5.15 eV/bond). To test the effect of photolysis, we placed 4 ml of TA solution in a quartz cuvette and exposed it to light in the 200–2500 nm wavelength range for 30 minutes, using a light source (Ocean Insight, DH-mini). However, no HTA was detected in the light-treated TA solution, indicating that UV light above 200 nm has a negligible influence on ·OH generation. Note that the study of Park et al. [30] identified nitrogen species (HONO) as a key species in UV-induced ·OH generation. The likelihood of HONO formation is very low in our experiment since our experiments were conducted in an argon atmosphere and under basic conditions. To further verify this, we added 50 $\mu$M NaNO$_2$ to the TA solution and exposed it to UV light for 5 minutes, but no HTA was detected. Additionally, electron collisions with water molecules can also produce ·OH. However, at energies below 9 eV, the collision cross-section is effectively zero [31], which significantly limits the likelihood of ·OH generation via electron collision processes.

In the previous study [20], ·OH generation was observed most abundantly in the order of electrolyte cathode, electrolyte anode, and zero-bias cases. In our study, for the case of the electrolyte anode, electrons shower onto the liquid surface without a significant voltage drop [27]. ·OH and e$_{aq}$ are highly concentrated just beneath the liquid surface, distributed with a shallow penetration depth and high concentration. Under these conditions, as the voltage increases, the plasma reaction area expands while the current density remains relatively constant. As a result, the diffused ·OH and e$_{aq}$ predominantly react at the surface, proportional to the increased reaction area. For the electrolyte cathode case, however, a large amount of ·OH is generated, most of which leads to H$_2$O$_2$ formation. Measuring e$_{aq}$ in this case is challenging due to surface instability caused by ion bombardment, and the mechanism of ·OH generation via ion bombardment requires further investigation.

## CONCLUSION

In the present study, we have investigated the concentration of e$_{aq}$ and ·OH in the underwater environment of plasma-liquid systems with the control of the bias voltage. To generate e$_{aq}$, a DC pulse discharge was applied using a needle and an electrolyte anode. The concentrations and penetration depth of e$_{aq}$ were measured using the TIRAS method, while ·OH concentrations were quantified using chemical dosimetry with a TA. The absorption profile of e$_{aq}$ undergoes notable changes with the bias voltage. Initially, as the voltage increases, the e$_{aq}$ exhibits an expanded profile, correlating with an increase in plasma radius. As the voltage continues to rise, the plasma profile extends beyond the laser spot, but the uniform pattern remains relatively stable even if it deviates from the Gaussian distribution. In the higher voltage range, the solvation pattern transforms from a uniform distribution to a ring pattern, causing a decrease in the laser beam absorption due to lower current density in the central area. These observations suggest a dynamic interaction between the bias voltage and the spatial distribution of the e$_{aq}$, ultimately affecting the absorbance measured by the TIRAS method. Across the voltage range of 1000–2500 V, the average penetration depth of eaq remained around 10 nm, with a peak concentration of approximately 1 mM. Diffusion is the primary mechanism for ·OH generation, with ·OH showing a distribution similar to e$_{aq}$ and primarily reacting with e$_{aq}$ at the interface. In contrast, at the electrolyte cathode, ·OH generation was significantly higher than could be explained by diffusion alone, resulting in high levels of hydrogen peroxide due to ·OH recombination.

**Acknowledgments**
This work was supported by the BK21 FOUR Program of the National Research Foundation of Korea (NRF) and by the Basic Science Research Institute (BSRI), POSTECH under the NRF grant No. 2021R1A6A1A10042944. This work was also supported by the Ministry of Trade, Industry & Energy (KEIT grant No. 20026366). The authors thank Dr. Suk-Ho Ahn at Pohang Accelerator Laboratory (PAL), POSTECH for his technical advice on the design and fabrication of the high voltage circuit used in the experiment.

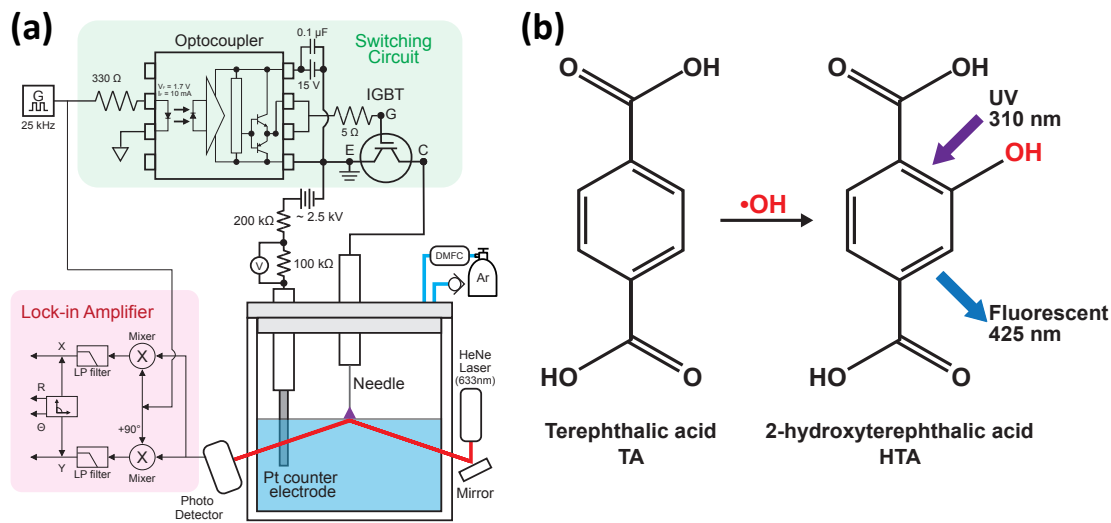

**Figure 1.** Schematic illustrations of (a) an experimental setup consisted of switching circuit and lock-in amplifier and (b) formation of HTA ($\lambda_{\text{excitation}}/\lambda_{\text{emission}} = 310/425\ nm$) through the reaction of TA and ·OH.

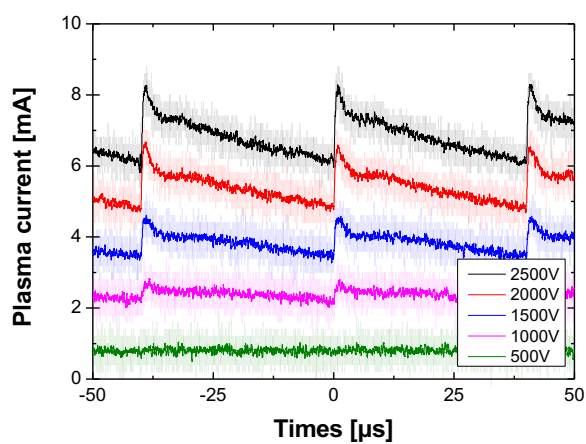

**Figure 2.** Waveforms of the plasma current at different bias voltages. The plasma current is modulated by a sawtooth wave.

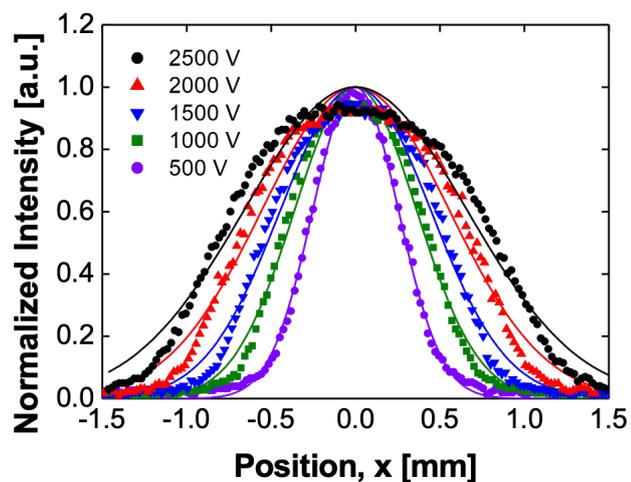

**Figure 3.** Plasma emission intensity profiles. This profile is obtained from the brightness information of the visible images taken with a DSLR camera. The plasma radius ($\sigma_p$) is obtained through Gaussian fitting.

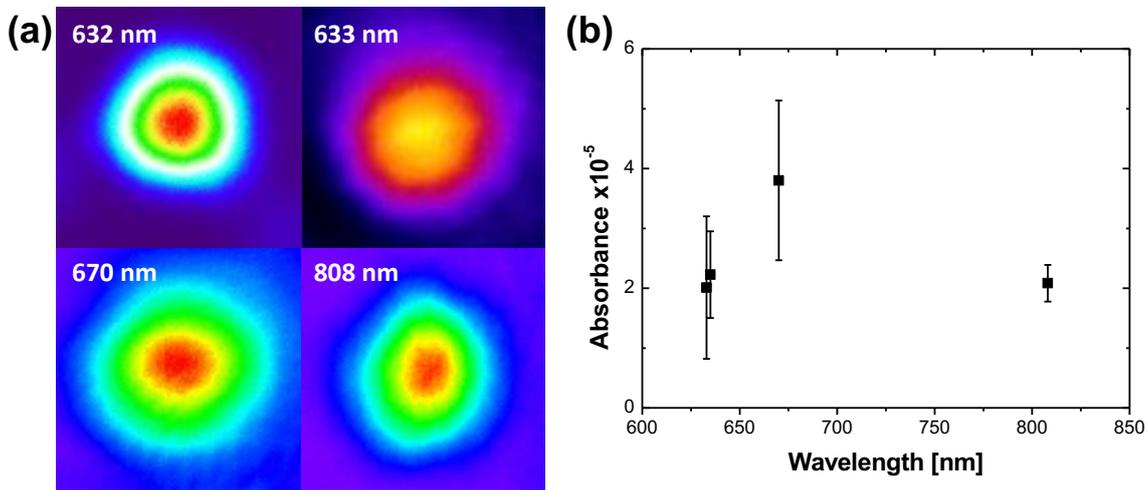

**Figure 4.** (a) The beam profile measured by a CCD. (b) Absorbance using HeNe laser (633 nm) and diode laser of different wavelengths (Thorlabs Inc.) at the 1500 V bias voltage

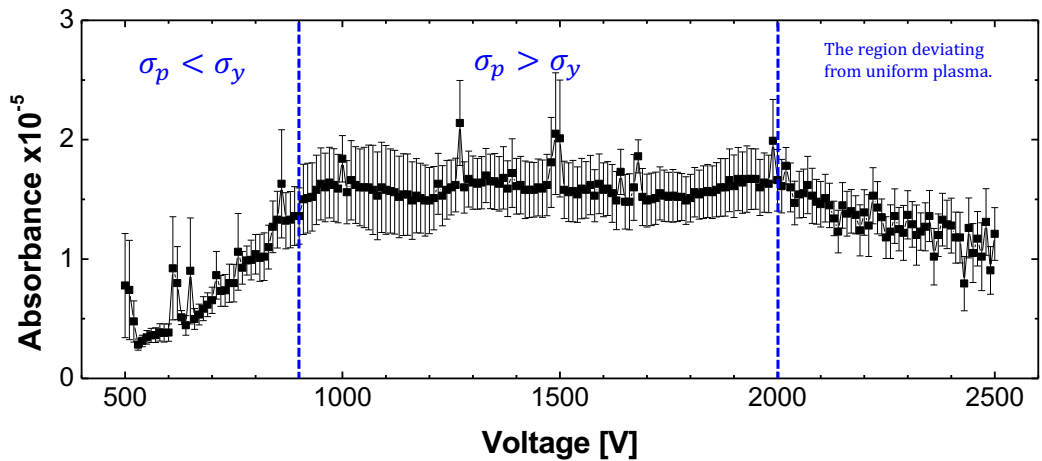

**Figure 5.** Optical absorbance due to solvated electrons at 633 nm as a function of the bias voltage.

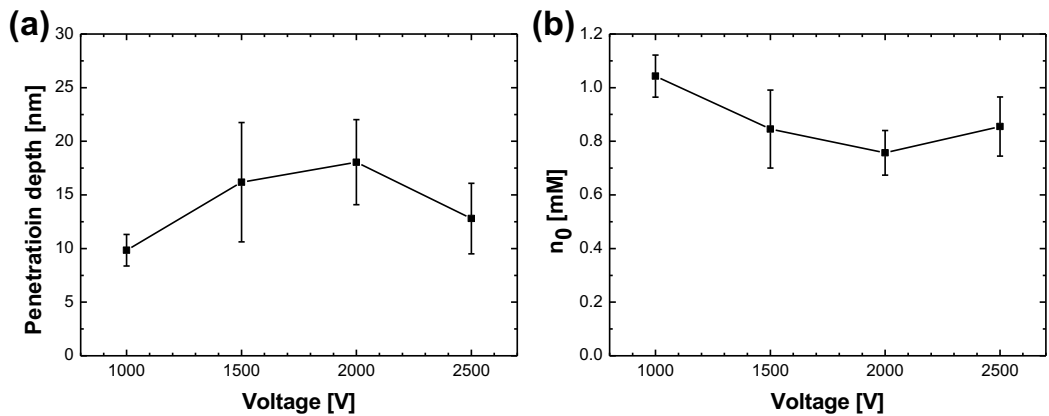

**Figure 6.** (a) Penetration depth and (b) central concentration of solvated electrons as a function of bias voltage.

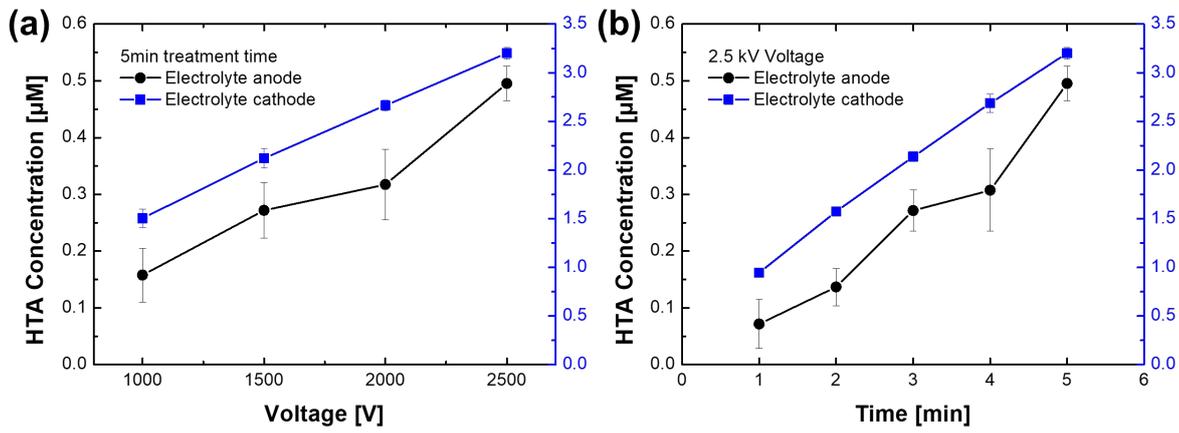

**Figure 7.** HTA concentration by varying bias voltage. Each sample consisting of 30 ml of TA solution was treated (a) for 1000 to 2500 V (b) for 5 min.

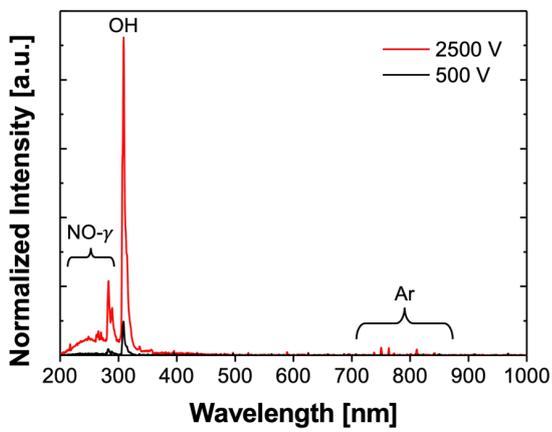

**Figure 8.** UV-Visible emission spectra of Ar plasma.